\documentstyle[sprocl1,epsfig]{article}

\def\be{\begin{equation}}
\def\ee{\end{equation}}
\def\bea{\begin{eqnarray}}
\def\eea{\end{eqnarray}}

\begin{document}

\title{Confinement and bound states in QCD}

\author{G. M. Prosperi}

\address{Dipartimento di Fisica dell'Universit\`a di Milano, via Celoria 16,
I20133 Milano, \\
I.N.F.N. Sez. di Milano, Italy. E-mail: prosperi@mi.infn.it}


\maketitle

\abstracts{I revue the so called Wilson loop approach to bound state problem
in QCD. I shall show how using appropriate path integral representations
for the quark propagator in an external field it is possible to obtain
corresponding path integral representations for various types of gauge
invariant Green functions which have the important feature 
of involving the gauge field only trough Wilson loop correlators or their
generalizations. Two different kinds of representations are used,
one given in the form of a semi-relativistic expansion, the second
completely relativistic of the Feynmann-Schwinger type. In this way starting
from reasonable ansatz on the non perturbative part of the Wilson correlator
one can obtain: expressions for the semi relativistic (spin dependent and 
momentum dependent)  $q \bar q$ and $3q$
potentials, a ``second order'' $q \bar q$ Bethe-Salpeter equation and
and a related Dyson-Schwinger equation. I shall concentrate on the three quark
potential for which new controversial results have been obtained by lattice
numerical simulations and on a three dimensional reduction of the BS equation
obtained in the form of the eigenvalue equation of of a squared or a usual
mass operator. We shall report on a numerical resolution of such equations
which seems to give a comprehensive reproduction of the entire meson spectrum 
with the exception of light pseudo-scalar bound states for which a complete 
four dimensional treatment should be necessary.}


\section{Green Functions}
     The single quark, the quark-antiquark and the three quark gauge 
invariant Green functions can be written as 
\begin{equation}
G^{\rm gi}(x-y) =
\langle0|{\rm T} U(y,x) \psi (x)
\overline{\psi}(y)  \overline{\psi} (y)
|0\rangle =
{\rm Tr_C} \langle U(y,x)
 S(x,y;A) \rangle
\label{eq:qgauginv} \, ,
\end{equation}
\begin{eqnarray}
G^{\rm gi}(x_1,x_2;y_1,y_2) &=&
\frac{1}{3}\langle0|{\rm T}\psi_2^c(x_2)U(x_2,x_1)\psi_1(x_1)
\overline{\psi}_1(y_1)U(y_1,y_2)  \overline{\psi}_2^c(y_2)
|0\rangle =
\nonumber\\
&=& \frac{1}{3} {\rm Tr_C} \langle U(x_2,x_1)
 S_1(x_1,y_1;A) U(y_1,y_2)
\tilde S_2(y_2,x_2;-\tilde A) \rangle
\label{eq:qqgauginv}
\end{eqnarray}
(the two quarks are supposed to have a different flavor, otherwise an 
annihilation term should be added to the last term)
\begin{eqnarray}
   && G^{\rm gi}(x_1,x_2,x_3;y_1,y_2,y_3)
    = \frac{1}{3!} \varepsilon_{a_1 a_2 a_3}
    \varepsilon_{b_1 b_2 b_3}
    \langle 0| {\rm T} \,
    U^{a_3 c_3}(x_M,x_3) U^{a_2 c_2}(x_M,x_2)
    \nonumber\\
   && U^{a_1 c_1}(x_M,x_1)
    \psi_{3 c_3}(x_3)
    \psi_{2 c_2}(x_2) \psi_{1 c_1}(x_1)
    \overline{\psi}_{1 d_1}(y_1) \overline{\psi}_{2 d_2}(y_2)
    \overline{\psi}_{3 d_3}(y_3)
    \nonumber\\
   && U^{d_1 b_1}(y_1,y_M) U^{d_2 b_2}(y_2,y_M)
    U^{d_3 b_3}(y_3,y_M) |0 \rangle = 
     \frac{1}{3!} \varepsilon_{a_1 a_2 a_3}
     \varepsilon_{b_1 b_2 b_3} 
     \nonumber \\
    && \big\langle \big( U(x_M,x_1)
     S_1(x_1,y_1;A) U(y_1,y_M) \big)^{a_1 b_1}
     \big( U(x_M,x_2) S_2(x_2,y_2;A)
     \nonumber\\
    && U(y_2,y_M)  \big)^{a_2 b_2}
     \big( U(x_M,x_3) S_3(x_3,y_3;A)
     U(y_3,y_M) \big)^{a_3 b_3}
     \big\rangle \> .
     \label{eq:treqgauginv}
     \end{eqnarray}
To avoid complicate indices in the above equations we have identified the 
various type of functions simply by their arguments; furthermore $\psi^c$ 
denotes the charge-conjugate fields, and 
$U$ the path-ordered gauge string (Schwinger string)
\begin{equation}
U(b,a)= {\rm P}  \exp \bigg\{ ig\int_a^b dx^{\mu} \, A_{\mu}(x) 
\bigg\} \, ,
\label{eq:col}
\end{equation}
(the integration in (\ref{eq:col}) is along an arbitrary line joining
$a$ to $b$), the tilde and ${\rm Tr_C}$ denote the transposition
and the trace over the color indices alone. The second equalities
are the result of an explicit integration over the fermionic fields  and the 
angle brackets denote average on the gauge variable alone (weighted in 
principle with the determinant  $M_f(A)$ resulting from the integration).
The quantities $S$, \ldots  $S_3$ are the quark propagators in the external 
gauge field $A^{\mu}$, which are defined by equations of the type
(we shall suppress indices specifying the quarks, as 
a rule, when dealing with single quark quantities)
\begin{equation}
( i\gamma^\mu D_\mu -m) S(x,y;A) =\delta^4(x-y)\,.
\label{eq:propdir}
\end{equation}
Here $m$ is supposed to hold an infinitesimal negative imaginary
part ($m=m-i0$).


\section{Quark-antiquark potential}

By performing a Foldy--Wouthuysen transformation on Eq. (\ref{eq:propdir})
we can replace the the $4\times4$  Dirac type matrices $S(x,y;A)$ with a
$2\times2$ Pauli propagator $ K({\bf x},{\bf y};t_{\rm f},t_{\rm i}) $ 
satisfying a Schr\"odinger-like equation. Solving such equation by standard 
path-integral technique, we obtain
\begin{eqnarray}
& & K({\bf x}, {\bf y}; t_{\rm f}, t_{\rm i})=
\int_{{\bf z}_1(t_{\rm i})={\bf y}_1}^{{\bf z}_1(t_{\rm f})={\bf x}_1}
    {\cal D} {\bf z} {\cal D} {\bf p}
\nonumber \\
& &  \exp \{i\int_{t_{\rm i}}^{t_{\rm f}} dt\,
[{\bf p} \cdot \dot{{\bf z}} - m -
\frac{{\bf p}^2}{2m}+\frac{{\bf p}^4}{8m^3} ]\}
{\rm T_s \, P} \exp \{ig\int_{\Gamma} dz^{\mu} \,
A_{\mu}(z) + \nonumber \\
& & + \frac{ig}{m_j} \int_{{\Gamma}} dz^{\mu}
  (S_j^l \hat{F}_{l{\mu}}(z) -\frac{1}{2m_j}
S_j^l\varepsilon^{lkr}p^k F_{{\mu}r}(z)-
\frac{1}{8m} D^{\nu}F_{{\nu}{\mu}}(z) 
 ) \}\,.
\label{eq:path1}
\end{eqnarray}
Here $\Gamma$ is the quark world line connecting $({\bf y},t_{\rm i})$ with 
$({\bf x},t_{\rm f})$ over which the path integral acts, 
${\rm T_s}$ and ${\rm P}$ are time ordering prescriptions for the
 spin and the color matrices respectively. Furthermore, as usual 
$F^{\mu\nu}= \partial^{\mu} A^{\nu} - \partial^{\nu} A^{\mu}
+ ig[A^{\mu},A^{\nu}]$,
$\hat{F}^{\mu\nu}= \frac{1}{2} \varepsilon^{\mu\nu\rho\sigma}
F_{\rho\sigma}$ and
$D^{\nu}F_{\nu\mu}= \partial^{\nu}F_{\nu\mu}+
ig[A^{\nu},F_{\nu\mu}]\>$,
 $\varepsilon^{\mu\nu\rho\sigma}$ being the four-dimensional Ricci 
symbol.

  Replacing (\ref{eq:path1}) in (\ref{eq:qqgauginv}) we obtain 
a path integral representation for a Pauli type $q \bar q$ propagator
which can
be compared with the corresponding expression in potential theory
(see [\cite{bcp}] for details). Then one find that, in order the
interaction 
between the two particles can be described in terms of a potential,
a function
$ V^{q\overline{q}}({\bf z}_1,{\bf z}_2,{\bf p}_1,{\bf p}_2,{\bf S}_1,
{\bf S}_2)$ must exist such that op to the order $ {1\over m^2}$  
\begin{eqnarray}
   i \ln W_{q\overline{q}}  +
   i \sum_{j=1}^2 \frac{ig}{m_j} \int_{{\Gamma}_j} \! dx^{\mu}
   \bigg( S_j^l \, \langle\langle \hat{F}_{l\mu}(x)
   \rangle \rangle -\frac{1}{2m_j} S_j^l
   \varepsilon^{lkr} p_j^k \, \langle\langle
   F_{\mu r}(x) \rangle\rangle -
    \nonumber\\
   {} - \frac{1}{8m_j} \, \langle\langle
   D^{\nu} F_{\nu\mu}(x) \rangle\rangle \bigg)
   - \frac{1}{2} \sum_{j,j^{\prime}} \frac{ig^2}{m_jm_{j^{\prime}}}
   {\rm T_s} \int_{{\Gamma}_j} dx^{\mu} \, \int_{{\Gamma}_{j^{\prime}}}
   dx^{\prime\sigma} \, S_j^l \, S_{j^{\prime}}^k
       \nonumber\\
   \bigg( \, \langle\langle \hat{F}_{l \mu}(x)
   \hat{F}_{k \sigma}(x^{\prime})
   \rangle\rangle - \, \langle\langle \hat{F}_{l \mu}(x) \rangle\rangle
   \, \langle\langle \hat{F}_{k \sigma}(x^{\prime}) \rangle\rangle \bigg)
   +\dots =\int_{t_{\rm i}}^{t_{\rm f}} dt
   \, V^{q\overline{q}}\> ,
\label{eq:defpot}
\end{eqnarray}
with the notation
$     \langle\langle f[A] \rangle\rangle = \left\langle 
      W_{q\overline q } f[A] \right\rangle /
      \left\langle W_{q\overline q } \right\rangle $
and $W_{q\overline{q}}$ is the Wilson loop correlator defined by 
\begin{equation}
W_{q\overline{q}} = \frac{1}{3} \left\langle {\rm Tr_C P}
 \exp \left( ig \oint_{\Gamma_{q \overline q}} dx^{\mu} \
  A_{\mu}(x) \right) \right\rangle \> .
\label{eq:wl}
\end{equation}
In Eq. (\ref{eq:wl}) the integration loop $\Gamma_{q \overline q}$ is supposed 
to be 
made by the quark world line  $\Gamma_1$, the antiquark world line $\Gamma_2$
described in reverse direction and the two Schwinger strings that close the
curve 
and can be here taken equal time straight lines; the symbol ${\rm P}$ denotes
color ordering along $\Gamma_{q \overline q}$. 

     Notice that the quantities $ \langle \langle F_{\mu\nu}(z_j) \rangle 
\rangle $ and  $ \langle \langle F_{\mu\nu}(z_j) F_{\rho\sigma}(z_j^\prime)
\rangle \rangle $ can be expressed as functional derivatives of 
$ i \ln W_{q\overline{q}} $ and so the potential is determined in principle
by such quantity alone.

    Eq. (\ref{eq:defpot}) can be reelaborated in various way. By expanding 
$i\ln W_{q\overline{q}}$ and the spin dependent terms in $\dot z_j$ it is
possible to recast the resulting local coefficients in terms of static
Wilson loop correlators (straight quark world-lines parallel to the time axis) 
with insertion of field strengths. Such expressions are particularly suitable
for numerical simulation and to this aim have been successfully
used [\cite{bali}].

   To obtain analytical expressions today we have to relay on
models for the evaluation of the Wilson correlator. The simplest model is the 
so called minimal area law model (MAL model) and consists in writing 
$i\ln W_{q\overline{q}}$ as the sum of its perturbative, an area and possibly 
a perimeter term
\begin{equation}
   i\ln W_{q\overline{q}} = (i\ln W_{q\overline{q}})_{\rm pert} + \sigma 
   S_{\rm min} + {1 \over 2}C P
\label{eq:mal}
\end{equation}
where $ S_{\rm min} $ denotes the minimal surface enclosed by the loop
$ \Gamma_{q \overline q} $  and $ P $  length of the loop.
In Eq. (\ref{eq:mal})
the first quantity is supposed to give correctly the short range limit
of the interaction and it is suggested by asymptotic freedom;
the other two are supposed to represent
the long range behavior and are suggested by pure lattice gauge 
theory and numerical simulations [\cite{wilson}].

         At the lowest order the perturbative term can be written as
\begin{equation}
i (\ln W_{q \overline q})_{\rm pert} = -{2 \over 3} g^2 \oint dz^\mu \oint
 dz^{\nu \prime} D_{\mu \nu}(z-z^\prime)
\label {eq:perta}
\end{equation}
$ D_{\mu \nu}(z-z^\prime) $ being the free gauge propagator.
For what concerns the area term it can be checked, by solving the appropriate 
Euler equations, that $ S_{\rm min} $ can be replaced by the surface
spanned by the straight lines joining equal time points on $\Gamma_1$ and 
$\Gamma_2$ up to ${\rm O}(v^2)$.
That is, we can write
\begin{equation}
 S_{\rm min} \approx 
\int_{t_{\rm i}}^{t_{\rm f}} dt \ \sigma r \int_0^1 d\lambda \ [1-(\lambda
\dot{{\bf z}}_{1 \rm T} + (1-\lambda)
 \dot{{\bf z}}_{2 \rm T} )^2]^{\frac{1}{2}}
\label{eq:sta}
\end{equation}
%
%
where $ \dot{\bf z}_{j {\rm T}}$ denotes the {\it transversal part}
 of $\dot{\bf z}_j $, 
$ \dot{z}_{j{\rm T}}^h=  (\delta^{hk} -\hat{r}^h \hat{r}^k)
 \dot{z}_j^h $.

   Replacing (\ref{eq:perta}) and (\ref{eq:sta}) in (\ref{eq:mal}) and 
(\ref{eq:defpot}), by expanding in the velocities and other manipulations
we obtain a semi-relativistic potential of the form $ V^{q \overline q}=
 V_{\rm static}^{q \overline q}+ V_{\rm sd}^{q \overline q}+ 
V_{\rm vd}^{q \overline q} $, where $ V_{\rm static}^{q \overline q}=
-{4 \over 3} \alpha_{\rm s}{1 \over r}+ \sigma r + C $, 
$ V_{\rm sd}^{q \overline q}$ and $ V_{\rm vd}^{q \overline q}$ are certain 
complicate spin dependent and velocity dependent expressions which are 
reported in [\cite{bcp}] but we do not reproduce here. Similar results can 
be obtained starting from more elaborate models [\cite{baker}].


\section{Three quark potential}

   A semi-relativistic three quark potential can be obtained
proceeding in a similar way on the three quark Green function
(\ref{eq:treqgauginv}). Instead of
(\ref{eq:defpot}) this time we have
\begin{eqnarray}
&& i \ln W_{3q}  + 
i \sum_{j=1}^3 \frac{ig}{m_j} \int_{{\Gamma}_j}dx^{\mu} 
\bigg( S_j^l \, \langle\langle \hat{F}_{l\mu}(x)
 \rangle \rangle -\frac{1}{2m_j} S_j^l
 \varepsilon^{lkr} p_j^k \, \langle\langle 
F_{\mu r}(x) \rangle\rangle -
\nonumber\\
&& - \frac{1}{8m_j} \, \langle\langle
D^{\nu} F_{\nu\mu}(x) \rangle\rangle \bigg)
 - \frac{1}{2} \sum_{j,j^{\prime}} \frac{ig^2}{m_jm_{j^{\prime}}}
{\rm T_s} \int_{{\Gamma}_j} dx^{\mu} 
\, \int_{{\Gamma}_{j^{\prime}}} 
dx^{\prime\sigma}
\, S_j^l \, S_{j^{\prime}}^k 
\nonumber \\
&& \bigg( \, \langle\langle \hat{F}_{l \mu}(x)
 \hat{F}_{k \sigma}(x^{\prime})
\rangle\rangle - \, \langle\langle \hat{F}_{l \mu}(x) \rangle\rangle
\, \langle\langle \hat{F}_{k \sigma}(x^{\prime}) \rangle\rangle \bigg)
 =  \int_{t_{\rm i}}^{t_{\rm f}} dt
 \, V^{3q}
\label{eq:3qpdef}
\end{eqnarray}
where now
$ \langle\langle f[A] \rangle\rangle= \langle
W_{3q} f[A] \rangle / \langle W_{3q} \rangle $
and $W_{3q}$ is the Wilson loop correlator for three quarks
\begin{eqnarray}
  W_{3q} = \frac{1}{3!}  \varepsilon_{a_1 a_2 a_3}
 \varepsilon_{b_1 b_2 b_3}  \left[ {\rm  P} \exp \left( ig
 \int_{\overline{\Gamma}_1} dx^{\mu_1} A_{\mu_1}(x) \right) 
\right]^{a_1 b_1}
\nonumber\\
\left[ {\rm P} \exp \left( ig \int_{\overline{\Gamma}_2}
 dx^{\mu_2} A_{\mu_2}(x) \right) \right]^{a_2 b_2}
  \left[  {\rm P} \exp \left( ig \int_{\overline{\Gamma}_3 } dx^{\mu_3}
A_{\mu_3}(x) \right) \right]^{a_3 b_3} .
\label{eq:w3}
\end{eqnarray}

   In (\ref{eq:w3}) $a_j,b_j$ are color indices, $j=1,2,3$ and 
$\overline{\Gamma}_j$  denote the  curve made by the world lines $\Gamma_j$ 
for the  quark $j$ joining $y_j$ to $x_j$, a straight line on the surface 
$t=t_{\rm i}$ merging from an arbitrary fixed point
$y_M$ and arriving to $y_j$,
another  straight line on the surface
$t=t_{\rm f}$ connecting the world line to a second fixed $x_M$.

   A straightforward generalization of the arguments used in the
quark-antiquark case
suggests to write in place of (\ref{eq:mal}) and ({\ref{eq:perta}) 
\begin{equation}
 i \ln W_{3q} = \frac{2}{3} g^2 \sum _{i<j}
 \int _{\Gamma _i} dx^{\mu}
_i \int _{\Gamma _j} dx^{\nu}_j \  i D_{\mu \nu} (x_i - x_j)
+ \sigma S_{\min} + {1\over 3} C P \, ,
\label{eq:3mal}
\end{equation}
where the perturbative term is taken at the lowest order in $\alpha_s$
and now $S_{\min}$ denotes the minimum among all the surfaces made by
three sheets
having the curves $\overline{\Gamma}_1$, $\overline{\Gamma}_2$ and 
$\overline{\Gamma}_3$ as contours and joining on a line $\Gamma_M$
connecting $y_{\rm M}$ with $x_{\rm M}$
(the minimum is understood at fixed $\bar{\Gamma}_j$
as the surfaces and $\Gamma_{\rm M}$ change ).
Obviously, $P$ denotes the total length of $\overline{\Gamma}_1$,
$\overline{\Gamma}_2$ and $\overline{\Gamma}_3$.

   As in the $q \overline q$ case, up to the second order in the velocities
$ S_{\min} $ can be replaced by the surface spanned by the straight lines
joining
the point $ {\bf z}_{M} (t)$ at the time $t$ with the equal time positions 
of the three quarks ${\bf z}_1 (t)$, $ {\bf z}_2 (t)$ and
${\bf z}_3 (t)$, $ {\bf z}_{M} (t)$ being constructed according to the 
following rule: if no angle in the
triangle made by ${\bf z}_1 (t)$, $ {\bf z}_2 (t)$ and
${\bf z}_3 (t)$ exceeds $120^0$ (configuration I), ${\bf z}_M (t)$
coincides with the point inside the triangle which sees  the three sides
under the same angle $120^0$; if one of the three angles in the 
triangle  is $\ge 120^0$ (configuration II), ${\bf z}_{M} (t)$ 
coincides with the corresponding vertex, let us say ${\bf z}_{\bar{j}}(t)$.

  The final result is of the 
form $ V^{3q} = V^{3q}_{{\rm stat}} + V^{3q}_{{\rm sd}} + V^{3q}_{{\rm vd}}$, 
where $ V^{3q}_{{\rm stat}}  = \sum_{j<l} \left( -\frac{2}{3}
\frac{\alpha_s}{r_{jl}} \right) + \sigma (r_1+r_2+r_3)+ C $
(with $ {\bf r}_j ={\bf z}_j -{\bf z}_{M}$,  ${\bf r}_{jl} = 
{\bf r}_j -{\bf r}_l \equiv {\bf z}_j - {\bf z}_l$), while
$V^{3q}_{{\rm sd}}$ and
$V^{3q}_{{\rm vd}}$ are spin dependent and momentum depend expressions of 
order $1/m^2$ for which again I refer to [\cite{bcp}].

  We observe also that the short range part in $V^{3q}$, coming from the first 
term in (\ref{eq:3mal}) is of  a  pure two body type: in fact it is identical
to the electromagnetic potential among three equally charged particles but
for the color group factor $2/3$.

  The three sheet definition of $S_{\rm min}$ given above 
is suggested by a direct generalization of Wilson's original argument for
$W_{q \bar q}$, based on the lattice formulation. It is adopted by the majority 
of the authors and it is said to correspond to an $Y$ flux tube configuration.
 However an alternative possibility has also been considered 
[\cite{cornwall}]. This consists in setting 
\begin{equation}
 S_{\rm min} = {1\over 2} \sigma (S_{12}+ S_{23}+S_{31})  \,,
\label{eq:deltaconf}
\end{equation}
where now $S_{ij}$ denotes the minimal surface delimited $\overline \Gamma_i $
and $\overline \Gamma_j$ and the factor ${1 \over 2}$ is introduced in order
to reproduce the $q \bar q$ case when two quark world-lines collapse.
Assumption (\ref{eq:deltaconf}) is said to correspond to a $\Delta$ 
flux tube configuration. In this case even the long range part of the 
three quark potential is the sum of purely two body potential and one can 
write simply $ V_{3q}= {1 \over 2} (V_{12}^{q \bar q}+
V_{23}^{q \bar q}+V_{31}^{q \bar q}), $ $V_{ij}^{q \bar q}$ being equal to
the quark antiquark potential relative to the couple $ij$ as
defined by (\ref{eq:defpot}).

If we restrict ourself to the linear rising part (the confining part)
of the static potentials alone we can write for the two configurations
\begin{eqnarray}
& & V_Y^{\rm conf} = \sigma (r_1+r_2+r_3) \label{eq:ypot}\,, \\
& & V_\Delta^{\rm conf} = {1 \over 2} \sigma (r_{12}+r_{23}+r_{31}) \,. 
\label{eq:deltapot}
\end{eqnarray}
Then, simple geometrical considerations show that
\begin{equation}
   V_Y^{\rm conf} \leq V_\Delta^{\rm conf} 
              \leq {2 \over \sqrt{3}}  V_Y^{\rm conf}  \,, 
\label{eq:ineq}
\end{equation}
where the equal sign in the first step holds when the three quarks are aligned,
in the second one when they form an equilateral triangle; the second one being 
the situation in which the difference between the two expressions is maximal 
(note, however that $ {2 \over \sqrt{3}} \sim 1.155 $). Until recently the 
various attempts to discriminate the two expressions by evaluating numerically
$\ln W_{3q}$ for a static loop have been unsuccessful. Presently preliminary 
results obtained by Bali et al. [\cite{bali}] for the equilateral arrangement 
are claimed to support $\Delta$ configuration. On the contrary the very 
accurate estimates reported by Matsufuru at this conference [\cite{matsufuru}]
are definitely in favour of the $Y$ configuration.
Indeed it seems that, due to the large errors occurring
in the results reported in Ref. [\cite{bali}}] at the 
large distances, these can be reasonably reconciled with those of Ref. 
[\cite{matsufuru}] simply by an appropriate choice of the constant $C$ 
(private communication by Suganuma). If this were confirmed, the $Y$
configuration should be considered established.


\section{Bethe-Salpeter and Dyson-Schwinger equation}

Full relativistic bound state equations can be obtained along similar lines
using a covariant representation for the solution of (\ref{eq:propdir}).

   To this aim it is convenient to rewrite the ``first order'' 
propagator $S(x,y;A)$ in terms of a ``second order'' one 
\begin{equation} 
S(x,y;A) = (i \gamma^\nu D_\nu + m) \Delta^\sigma(x,y;A) \,,
\label{eq:fsord}
\end{equation}
$\Delta^\sigma(x,y;A)$ being defined by 
\begin{equation}
(D_\mu D^\mu +m^2 -{1\over 2} g \, \sigma^{\mu \nu} F_{\mu \nu})
\Delta^\sigma (x,y;A) = -\delta^4(x-y) \, ,
\label{eq:propk}
\end{equation}
with $\sigma^{\mu \nu} = {i\over 2} [\gamma^\mu, \gamma^\nu]$.

    After replacing (\ref{eq:fsord}) in (\ref{eq:qqgauginv}), using an 
appropriate derivative it is possible to take the differential operator out 
of the angle brackets and write [\cite{prosp96}]
\begin{equation}
G^{\rm gi}(x_1,x_2; y_1,y_2) =-(i \gamma_1^\mu \bar{\partial}_{1 \mu} 
 + m_1) ( i \gamma_2^\nu \bar{\partial}_{2 \nu} +m_2) 
H^{\rm gi}(x_1,x_2;y_1,y_2) \, ;
\label{eq:qqeqg}
\end{equation}
a similar expression can be given for $G^{\rm gi}(x-y)$, having set
\begin{eqnarray}
H^{\rm gi}(x_1,x_2;y_1,y_2)&=& -{1\over 3} {\rm Tr _C}
\langle U(x_2,x_1) \Delta_1^\sigma (x_1,y_1;A)
 U(y_1,y_2) \tilde{\Delta}_2^\sigma (x_2,y_2;-\tilde{A})\rangle \, ,
\nonumber \\
H^{\rm gi}(x-y)&=& i {\rm Tr _C}
\langle U(y,x) \Delta ^\sigma (x,y;A) \rangle .
\label{eq:qqeqh}
\end{eqnarray}

  For the second order propagator we have the Feynmann-Schwinger 
representation
\begin{eqnarray}
\Delta^\sigma (x,y; A) &=&  -{i \over 2} \int_0^\infty ds \int_y^x
   {\cal D} z \exp [-i \int_0^s d\tau {1\over 2} (m^2 
   +\dot z ^2)] \nonumber \\
  & & \qquad \qquad {\cal S}_0^s {\rm P} \exp[ ig \int_0^s d \tau
   \dot z ^\mu A_{\mu}(z)] \, , 
\label{eq:partbis}
\end{eqnarray}
where the world-line connecting $y$ to $x$ is specified in the
four-dimensional language by $z^\mu=z^\mu(\tau)$, in terms of
an additional parameter $\tau$, and
$ {\cal S}_0^s = {\rm T} \exp \Big [ -{1\over 4} \int_0^s d \tau 
 \sigma^{\mu \nu} {\delta \over \delta S^{\mu \nu}(z)}
\Big ] $ and $ \delta 
S^{\mu \nu} = dz^\mu \delta z^\nu - dz^\nu \delta z^\mu $ (the functional 
derivative being defined through an arbitrary deformation, $ z \rightarrow 
z + \delta z $, of the world-line). 

   Substituting (\ref{eq:partbis}) into (\ref{eq:qqeqh}), we obtain
covariant path integral representations for
$H^{\bf gi}(x_1,x_2;y_1,y_2)$ and $H^{\bf gi}(x-y)$.
These representations involve again the gauge field only trough the Wilson 
correlator $W_{q \overline q}$ for the four point function and a similar 
quantity $W_q$ for the two point function. $W_q$ is obtained
replacing in (\ref{eq:wl}) the loop $\Gamma_{q \overline q}$ by a second
loop made by the world-line connecting $y$ to $x$ closed by the appropriate
Schwinger string [\cite{prosp96}]. 
%
%
%

  For the Wilson loop correlator we can make the ansatz (\ref{eq:mal}
-\ref{eq:sta})in the center of mass system even in the relativistic case, 
after taking $C=0$ (since in a 
relativistic treatment the perimeter term can be completely reabsorbed in a 
mass renormalization). Notice that in such a case in general (\ref{eq:sta}) is 
only an approximation, however, if rewritten in an appropriate way,
it seems to be a significant one. In fact, beside to have the correct 
semi-relativistic limit,(\ref{eq:sta}) becomes exact in important geometrical 
situations [\cite{prosp96}] like those of a flat Wilson loop or of uniform 
rotatory motion. 

  From the above path integral representations and by an 
appropriate recurrence method, an inhomogeneous Bethe-Salpeter equation and a 
Dyson-Schwinger equation can be derived for two other second order
functions \\
$H(x_1,x_2,y_1,y_2)$ and $H(x-y)$, which are simply related to
$H^{\rm gi}(x_1,x_2,y_1,y_2)$ and $H^{\rm gi}(x-y)$, reduce to them 
in the limit of vanishing $x_1-x_2,\ y_1-y_2$ or $x-y$ and are completely 
equivalent for what concerns the determination of bound states, effective 
masses, etc. 

 In the momentum space, the corresponding homogeneous BS-equation can be 
written (in a $4 \times 4$ matrix representation)
\begin{eqnarray}
  \Phi_P (k) &=& -i \int \! {d^4u \over (2 \pi)^4} \;
        \hat I_{ab} \big (k-u, {1 \over 2}P +{k+u \over 2},
              {1 \over 2}P-{k+u \over 2} \big )\nonumber \\
   & & \qquad \qquad
     \hat H_1   \big ({1 \over 2} P  + k \big )\sigma^a  \Phi_P (u) \sigma^b
     \hat H_2 \big (-{1 \over 2} P + k \big ) \, ,
\label{eq:bshoma}
\end{eqnarray}
where $\Phi_P (k)$ denotes an appropriate wave function, $\sigma^0=1$, 
$a=0,\mu \nu$ and the center of mass frame has to be understood (i.e. $P=
(m_B, {\bf 0})$, $m_B$ being the bound state mass). 
Similarly, 
the DS-equation can be written also
\begin{equation}
\hat \Gamma(k) =  \int \! {d^4 l \over (2 \pi)^4}  \;
\hat I_{ab} \Big ( k-l;{k+l \over 2},{k+l \over 2} \Big )
\sigma^a \hat H(l) \, \sigma^b \ .
\label{eq:sdeq}
\end{equation}
$\hat \Gamma (k)$ being the irreducible self-energy, defined by
$\hat H(k) =\hat H_0(k) + i\hat H_0(k)\hat \Gamma (k) \hat H(k) $.

  Notice that in principle  (\ref{eq:bshoma}) and  (\ref{eq:sdeq}) are 
exact equations. However
the kernels $\hat I_{ab}$ are generated in the form of an expansion in
$\alpha_{\rm s}$ and the string tension $\sigma$. At the lowest order
in both such constants, we have explicitly
\begin{eqnarray}
& & \hat I_{0;0} (Q; p, p^\prime)  = 
       16 \pi {4 \over 3} \alpha_{\rm s} p^\alpha p^{\prime \beta}
        \hat D_{\alpha \beta} (Q)  + \nonumber \\
& &  + 4 \sigma  \int d^3 {\bf \zeta} e^{-i{\bf Q}
              \cdot {\bf \zeta}}
         \vert {\bf \zeta} \vert \epsilon (p_0) \epsilon (p_0^\prime )
         \int_0^1 d \lambda \{ p_0^2 p_0^{\prime 2} -
         [\lambda p_0^\prime {\bf p}_{\rm T} +
         (1-\lambda) p_0 {\bf p}_{\rm T}^\prime ]^2 \} ^{1 \over 2}
\nonumber \\
& & \hat I_{\mu \nu ; 0}(Q;p,p^\prime) = 4\pi i {4 \over 3} \alpha_{\rm s}
   (\delta_\mu^\alpha Q_\nu - \delta_\nu^\alpha Q_\mu) p_\beta^\prime
   \hat D_{\alpha \beta}(Q)  - \nonumber \\
& & \qquad \qquad \qquad  - \sigma  \int d^3 {\bf \zeta} \, e^{-i {\bf
      Q} \cdot \zeta} \epsilon (p_0) {\zeta_\mu p_\nu -\zeta_\nu p_\mu \over
      \vert {\bf \zeta} \vert \sqrt{p_0^2-{\bf p}_{\rm T}^2}}
      p_0^\prime  \nonumber \\
& & \hat I_{0; \rho \sigma}(Q;p,p^\prime) =
     -4 \pi i{4 \over 3} \alpha_{\rm s}
     p^\alpha (\delta_\rho^\beta Q_\sigma - \delta_\sigma^\beta Q_\rho)
     \hat D_{\alpha \beta}(Q) + \nonumber \\
& & \qquad \qquad  \qquad  + \sigma  \int d^3 {\bf
\zeta} \, e^{-i{\bf Q}
  \cdot {\bf \zeta}} p_0
  {\zeta_\rho p_\sigma^\prime - \zeta_\sigma p_\rho^\prime \over
    \vert {\bf \zeta} \vert \sqrt{p_0^{\prime 2}
   -{\bf p}_{\rm T}^{\prime 2}} }
    \epsilon (p_0^\prime)  \nonumber \\
& & \hat I_{\mu \nu ; \rho \sigma}(Q;p,p^\prime) =
    \pi {4\over 3} \alpha_{\rm s}
    (\delta_\mu^\alpha Q_\nu - \delta_\nu^\alpha Q_\mu)
    (\delta_\rho^\alpha Q_\sigma - \delta_\sigma^\alpha Q_\rho)
     \hat D_{\alpha \beta}(Q)
 \label{eq:imom}
 \end{eqnarray}
\noindent
where in the second and in the third equation $\zeta_0 = 0$ has to be
understood.


\section{Reduction of the BS equation and spectrum.}

   To  find the $q \overline q $ spectrum, in principle one should solve first 
(\ref{eq:sdeq}) and use the resulting propagator in (\ref{eq:bshoma}). In 
practice this turns out to be a difficult task and one has to resort to the 
three dimensional equation which can be obtained from (\ref{eq:bshoma}) by the 
so called instantaneous approximation. This consists
in replacing $\hat H_j(p)$ in (\ref{eq:bshoma}) with the free 
quark propagator ${-i \over p^2 -m_j^2}$ and the kernel 
$\hat{I}_{ab}$ with $ \hat{I}_{ab}^{\rm inst}({\bf k}, {\bf k}^\prime)$ 
obtained from $\hat{I}_{ab}$
setting $k_0=k_0^\prime= {m_2 \over m_1+m_2} { w_1+ w_1^\prime \over 2}-
{m_1 \over m_1+m_2} {w_2 +w_2^\prime \over 2}$ with $w_j = \sqrt{m_j^2 + 
{\bf k}^2} $ and $ w_j^\prime = \sqrt{m_j^2 + {\bf k}^{\prime 2}}$. 
%
%

    The reduced equation takes the form of the eigenvalue equation  for a 
squared mass operator,
$
    M^2 = M_0^2 + U \,,
$
with  $ M_0 = \sqrt{m_1^2 + {\bf k}^2} + \sqrt{m_2^2 + {\bf k}^2} $ and
\begin{equation}
   \langle {\bf k} \vert U \vert {\bf k}^\prime \rangle =
        {1\over (2 \pi)^3 }
        \sqrt{ w_1 + w_2 \over 2  w_1  w_2}\, \hat I_{ab}^{\rm inst}
        ({\bf k} , {\bf k}^\prime)  \sqrt{ w_1^\prime + w_2^\prime \over 2
         w_1^\prime w_2^\prime}\sigma_1^a \sigma_2^b \,.
\label{eq:quadrrel}
\end{equation}
The quadratic form of the above equation obviously derives from the second 
order character of the formalism we have used. 

  Alternatively, in more usual terms, one can look for the eigenvalue of the 
mass operator or center of mass hamiltonian 
$          H_{\rm CM} \equiv M = M_0 + V  $
with $V$ defined by $M_0V+VM_0+V^2=U$. Neglecting, consistently, the second 
order term $V$ can be obtained from $U$ simply by the kinematic replacement
$
\sqrt{ (w_1+w_2) (w_1^\prime +w_2^\prime)\over w_1w_2w_1^\prime w_2^\prime}
\to {1\over 2\sqrt{w_1 w_2 w_1^\prime w_2^\prime}}
$. 

   In ref. [\cite{mio}] we have evaluated the spectrum using both the operator
$M^2$ and $H_{\rm CM}$, including the hyperfine terms in the potentials but 
omitting the spin-orbit ones.
The numerical procedure we have followed was very simple.
It consisted in solving 
first the eigenvalue equation for the zero order hamiltonian $ H_0^{\rm CM}=
M_0 -{4\over 3}{\alpha_{\rm s} \over r}+\sigma r $ by
the Rayleigh-Ritz method, using the three-dimensional harmonic
oscillator basis and diagonalising a $30 \times 30$ matrix. Then we have
evaluated the quantities $\langle \psi_{\nu}|H_{\rm CM}|\psi_{\nu} \rangle$
and $\langle \psi_{\nu}|M^2|\psi_{\nu} \rangle$ for the eigenfunctions 
$\psi_{\nu}$ obtained in the first step. Since the quantity $\langle V^2 
\rangle$ was not completely negligible (ranging e. g. between few tens
and 150 MeV
for $c\bar c$), we had to use slight different values of the adjustable
parameters in the linear and in the quadratic formulation.
\begin{figure}[htbp!]
  \begin{center}
    \leavevmode
    \setlength{\unitlength}{1.0mm}
    \begin{picture}(140,70)
      \put(-15,0){\mbox{\epsfig{file=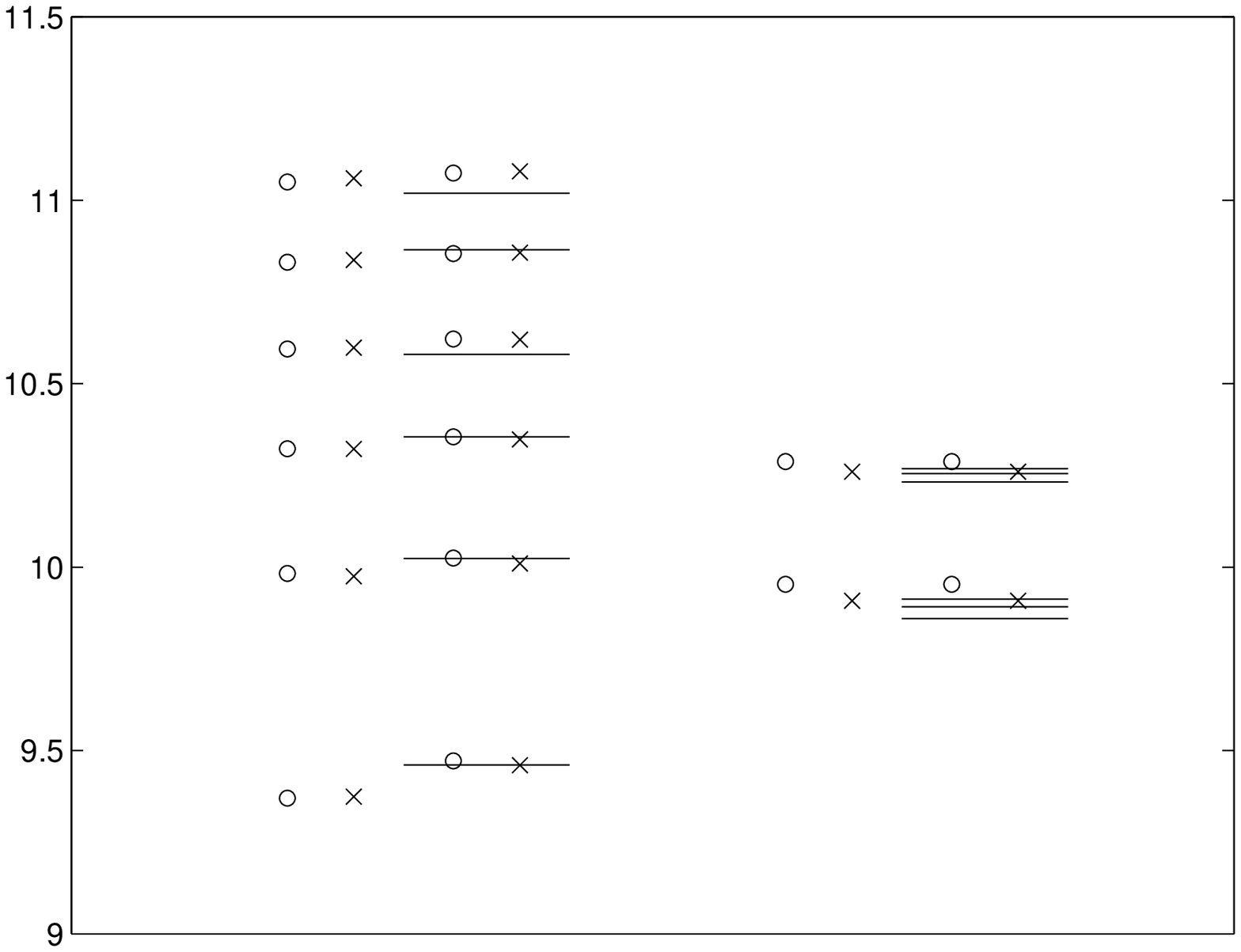,height=7cm,width=7.5cm}}}
      \put(41,6){ $ b \bar{b} $ }
      \put(0,61){ $ {^{1} {\rm S}_{0}} $ }
      \put(10,61){ $ {^{3} {\rm S}_{1}} $ }
      \put(30,61){ $ {^{1} {\rm P}_{1}} $ }
      \put(40,61){ $ {^{3} {\rm P}_{J}} $ }
      \put(20,61){ n }
      \put(20,55){ $ {}_{6} $ }
      \put(20,50){ $ {}_{5} $ }
      \put(20,43){ $ {}_{4} $ }
      \put(20,37){ $ {}_{3} $ }
      \put(20,28){ $ {}_{2} $ }
      \put(20,13){ $ {}_{1} $ }
      \put(50,61){ n }
      \put(50,35){ $ {}_{2} $ }
      \put(50,25){ $ {}_{1} $ }
      \put(70,0){\mbox{\epsfig{file=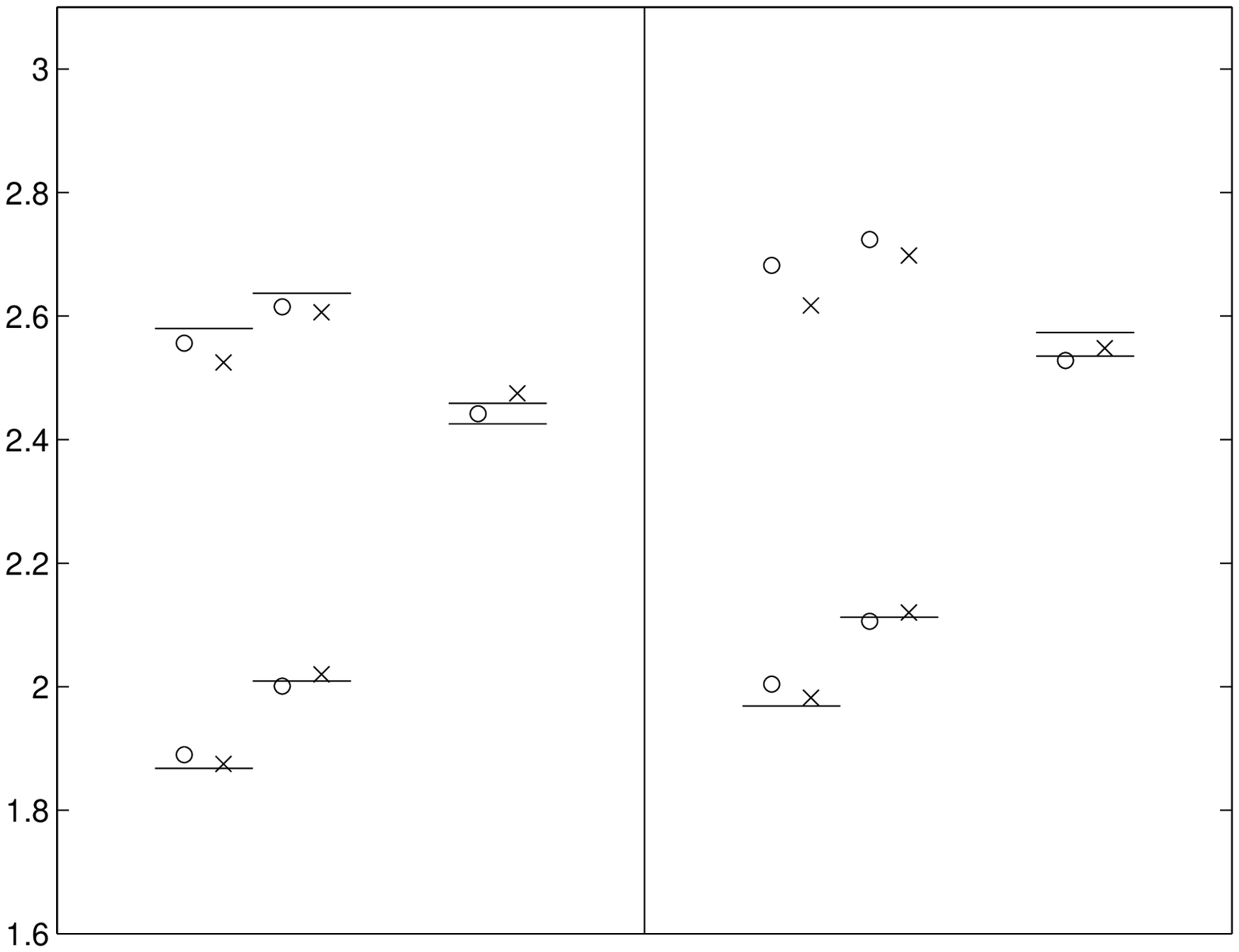,height=7cm,width=7.5cm}}}
      \put(97,6){ $ u \bar{c} $ }
      \put(133,6){ $ s \bar{c} $ }
      \put(76,61){ $ {^{1} {\rm S}_{0}} $ }
      \put(84,61){ $ {^{3} {\rm S}_{1}} $ }
      \put(98,61){ $ {\rm P} $ }
      \put(112,61){ $ {^{1} {\rm S}_{0}} $ }
      \put(120,61){ $ {^{3} {\rm S}_{1}} $ }
      \put(134,61){ $ {\rm P} $ }
    \end{picture}
  \end{center}
\caption{Heavy-heavy and light-heavy quarkonium spectra.}
\label{fig1}
\end{figure}
\begin{figure}[htbp!]
  \begin{center}
    \leavevmode
    \setlength{\unitlength}{1.0mm}
    \begin{picture}(140,70)
      \put(-11,0){\mbox{\epsfig{file=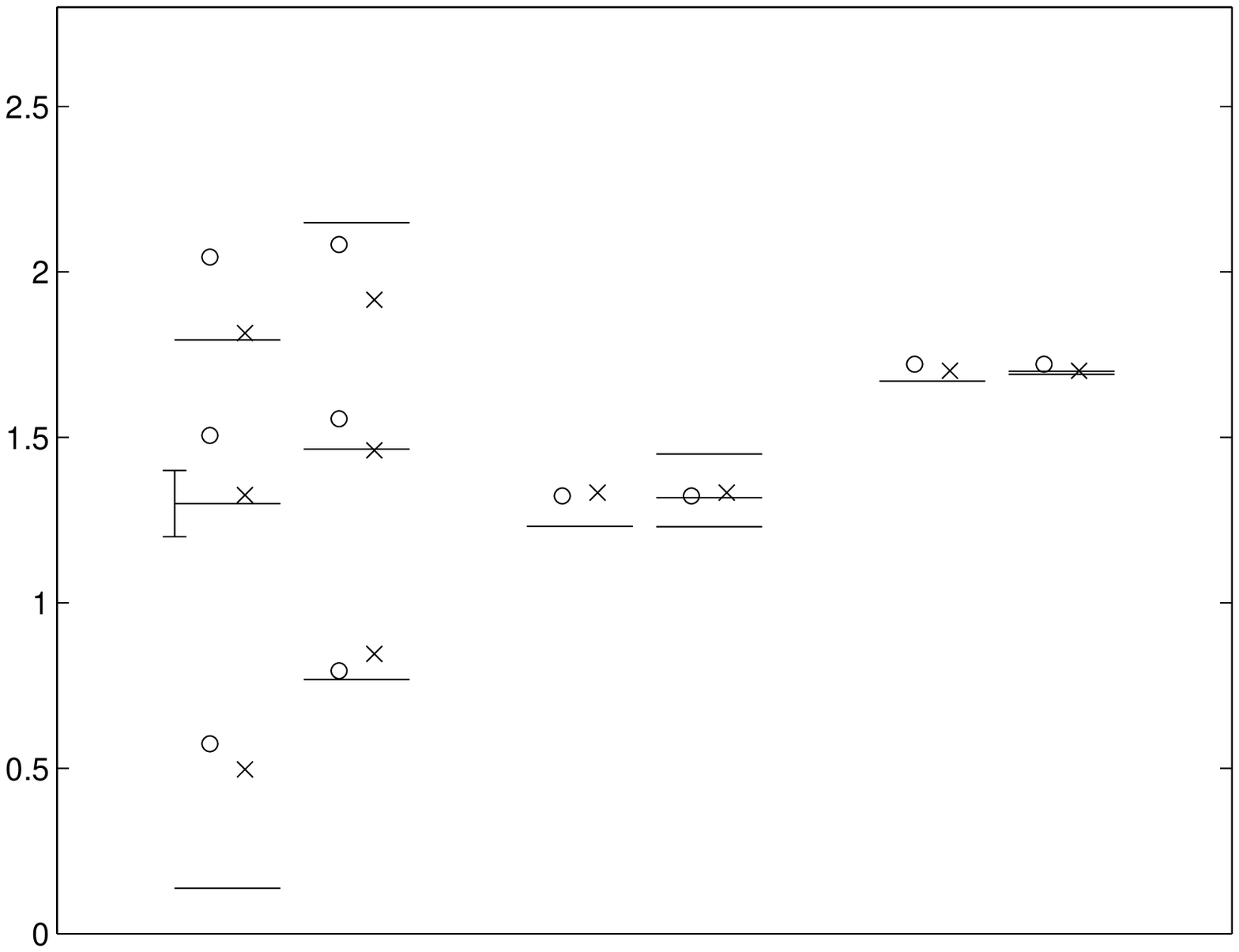,height=7cm,width=5.5cm}}}
      \put(30,5){ $ u \bar{u} $ }
      \put(-7,61){ $ {^{1} {\rm S}_{0}} $ }
      \put(0,61){ $ {^{3} {\rm S}_{1}} $ }
      \put(9,61){ $ {^{1} {\rm P}_{1}} $ }
      \put(16,61){ $ {^{3} {\rm P}_{J}} $ }
      \put(25,61){ $ {^{1} {\rm D}_{1}} $ }
      \put(32,61){ $ {^{3} {\rm D}_{J}} $ }
      \put(39,55){ $ {}_{J} $ }
      \put(37.5,42.5){ $ {}_{1,3} $ }
      \put(22,36.5){ $ {}_{?} $ }
      \put(70,0){\mbox{\epsfig{file=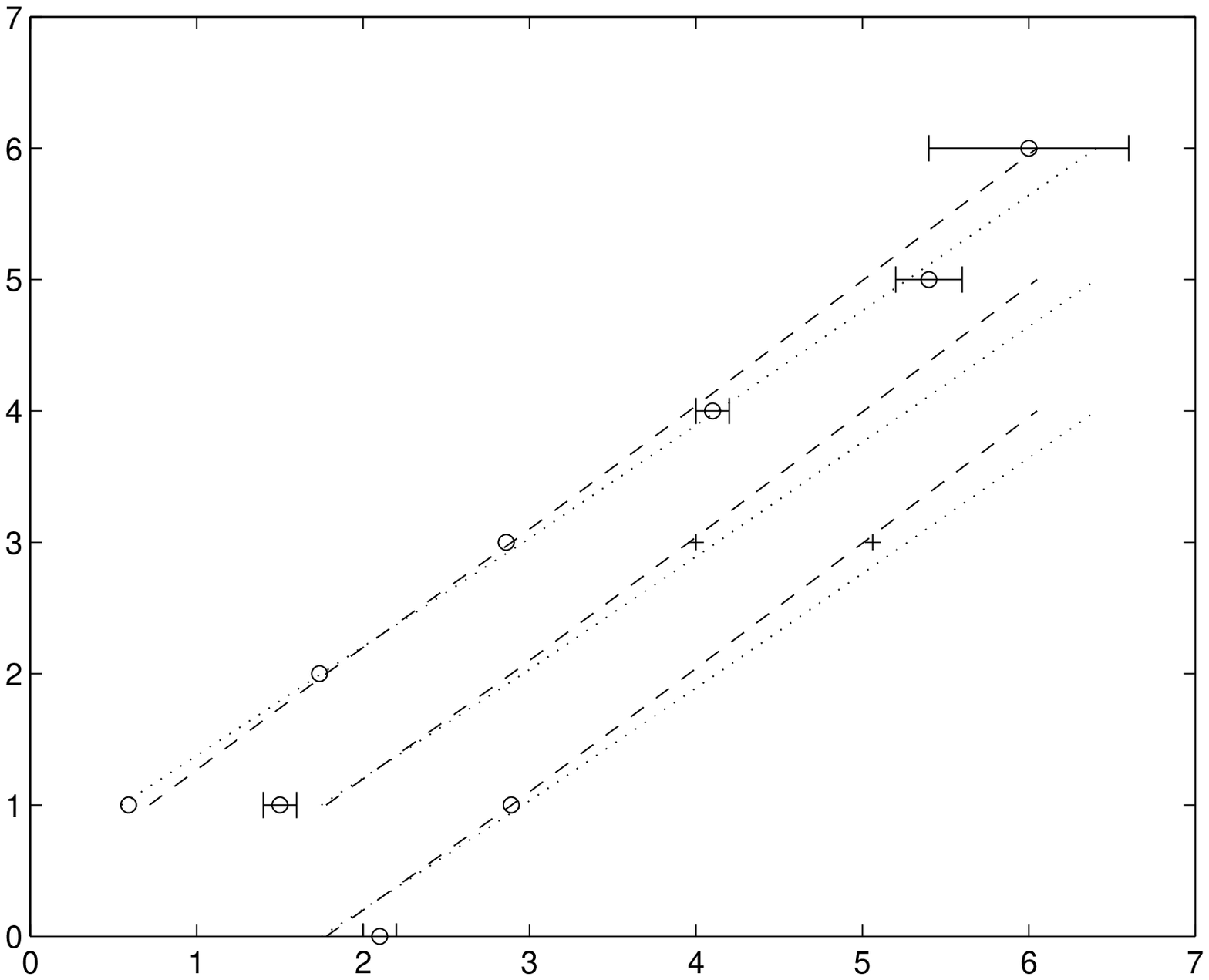,height=7cm,width=7cm}}}
      \put(80,60){ $ u \bar{u} $ }
      \put(66,63){ $ J $ }
      \put(140,1){ $ M^{2} $ }
      \put(128,61){ $ J = L + 1 $ }
      \put(128,51.5){ $ J = L $ }
      \put(128,42){ $ J = L - 1 $ }
      \put(68,15){ $ \rho ( 770 ) $ }
      \put(72,22){ $ a_{2} ( 1320 ) $ }
      \put(75,7){ $ a_{1} ( 1260 ) $ }
      \put(85,-3){ $ a_{0} ( 1450 ) $ }
      \put(83,31){ $ \rho_{3} ( 1690 ) $ }
      \put(100,11){ $ \rho ( 1700 ) $ }
      \put(93,40){ $ a_{4} ( 2040 ) $ }
      \put(100,26){ $ X ( 2000 ) $ }
      \put(105,50){ $ \rho_{5} ( 2350 ) $ }
      \put(121,28){ $ \rho_{3} ( 2250 ) $ }
      \put(112,61){ $ a_{6} ( 2450 ) $ }
    \end{picture}
  \end{center}
\caption{$ u \bar{u} $ spectrum and Regge trajectories.}
\label{fig2}
\end{figure}
We have kept the light quark masses fixed in both cases
on typical current values, $m_u=m_s=10\, {\rm MeV},\ m_s=200 \, {\rm MeV}$.
Then for the calculations based on $H_{\rm CM}$ we have assumed 
$m_c=1.40\, {\rm GeV}$, $m_b=4.81\, {\rm GeV}$,  $\alpha_{\rm s}=0.363$
and $\sigma = 0.175\, {\rm GeV}^2$, taken essentially from the
semi-relativistic fits. On the contrary for the calculations
based on $M^2$ we have used a running coupling constant of the form
\begin{equation}
\alpha_{\rm s}(\bf Q)= {4 \pi \over (11-{2 \over 3}N_{\rm f}) \ln {{\bf Q}^2
      \over \Lambda^2}}
      \label{eq:running}
      \end{equation}
cut at a maximum adjustable value $\alpha_{\rm s}(0)$ and with $N_{\rm f}=4$, 
$\Lambda=200\, {\rm MeV}$. Furthermore we have chosen $\alpha_{\rm s}(0)=0.35$ 
and $\sigma = 0.2\, 
{\rm GeV}^2$ in order to reproduce the correct $J/\Psi - \eta_c$ separation
and the Regge trajectory slope; $m_c=1.394\, {\rm GeV}$ and
$m_b=4.763\, {\rm GeV}$ in order to obtain exactly the masses of $J/\Psi$
and $\Upsilon$.

In both cases, on the whole, the agreement with the data is good, not only for
bottonium and charmonium (as in ordinary potential models), but also for the 
light-light and light-heavy systems which are here essentially parameter free.
Examples are reported in figs. 1 and 2. Circlet and dotted lines refer to the 
linear formulation, crosses and broken lines to the quadratic one. The only 
serious disagreement concerns the light pseudo-scalar mesons related to the 
chiral symmetry breaking problem. In fact
in this case, as well known, a strict connection should exist between wave 
function in (\ref{eq:bshoma}) and irreducible self energy in (\ref{eq:sdeq})
and the use of the free propagator can not be correct.


\end{document}